\documentclass[aps,prl,floatfix,twocolumn,showpacs,preprintnumbers,amsmath,amssymb,superscriptaddress]{revtex4-2}
\usepackage{amsmath}
\usepackage{amssymb}
\usepackage{graphicx}
\usepackage{hyperref} 
\usepackage{bm}
\usepackage{epstopdf}
\usepackage{epsfig}
\usepackage{comment}
\usepackage{color}

\begin{document}
\title{Polarization response of spin-lasers under amplitude modulation}
\author{Gaofeng Xu}
\email{xug@hdu.edu.cn}
\affiliation{Department of Physics, Hangzhou Dianzi University, Hangzhou, Zhejiang 310018, China}
\author{Krish Patel}
\affiliation{Department of Physics, University at Buffalo, State University of New York, Buffalo, NY 14260, USA}
\author{Igor \v{Z}uti\'c}
\email{zigor@buffalo.edu}
\affiliation{Department of Physics, University at Buffalo, State University of New York, Buffalo, NY 14260, USA}

\begin{abstract}
Lasers with injected spin-polarized carriers  show an outstanding performance in both static and dynamic operation. 
In addition to the intensity response of conventional lasers, without spin-polarized carriers, both intensity and polarization of light can be exploited for optical communication in spin-lasers. However, the polarization dynamics of spin-lasers under amplitude modulation has been largely overlooked. Here we reveal, 
analytically and numerically, a nontrivial polarization response that accompanies the well-known intensity dynamics of a spin-laser under amplitude modulation. 
We evaluate the polarization and intensity response under the same amplitude modulation, and further assess the capability of such a polarization response in digital data transfer with eye diagram simulations. Our results provide a more complete understanding of the modulation response in spin-lasers and open up unexplored opportunities in optical communication and spintronics.  
\end{abstract} 
\maketitle

Lasers are key devices in optical communication networks~\cite{Chuang:2009, Coldren:2012,Michalzik:2013, Ma2019:NN},
typically using the intensity  of the emitted light under amplitude modulation (AM)~\cite{Agrawal:2002}.  
However, injecting spin-polarized carriers into the lasers gives rise to new opportunities, which can improve their performance  
in both static and dynamic operation, including reduced lasing threshold and enhanced modulation  bandwidth~\cite{
Rudolph2003:APL,Rudolph2005:APL,Holub2007:PRL,Gerhardt2006:EL,Hovel2008:APL,Saha2010:PRB,Gerhardt2011:APL,Iba2011:APL,
Frougier2013:APL,Cheng2014:NN,Alharthi2014:APL,Alharthi2015:APL, Lindemann2019:N,Maksimov2022:JETPL}.
In such spin-lasers, through conservation of angular momentum, the spin polarization of carriers can be converted into the circular polarization of the emitted light, which enables control and modulation of the polarization of emission and leads to room-temperature spintronic applications beyond magnetoresistance~\cite{Zutic2004:RMP,Nishizawa2017:PNAS,Zutic2020:SSC,Rozhansky2021:PE,
Nishizawa2021:MM,Tsymbal:2019,Dery2011:APL,Khaetskii2013:PRL,Zutic2019:MT}. 

In a typical polarization modulation (PM) scheme, 
the spin-polarized injection can be modulated electrically or optically, which leads to a modulation of carrier spin polarization and the circularly polarized emitted light~\cite{Lee2010:APL, Yokota2018:APL,Lindemann2019:N, Drong2021:PRA,Huang2021:IEEEJQE}. 
For a spin-laser with large birefringence, the response of the circular polarization of light under polarization modulation has been shown to support a significantly enhanced modulation bandwidth and is promising for future ultrafast optical communication~\cite{FariaJunior2015:PRB,Lindemann2016:APL,Pusch2015:EL,Pusch2017:APL, Pusch2019:EL,  
Fordos2017:PRA,Yokota2021:MM,Drong2021:PRA,Huang2021:IEEEJQE,Heermeier2022:LPR,Lindemann2019:N,Tselios2022:PSSB}. 

\begin{figure}[ht]
\centering
\includegraphics*[width=8.6cm]{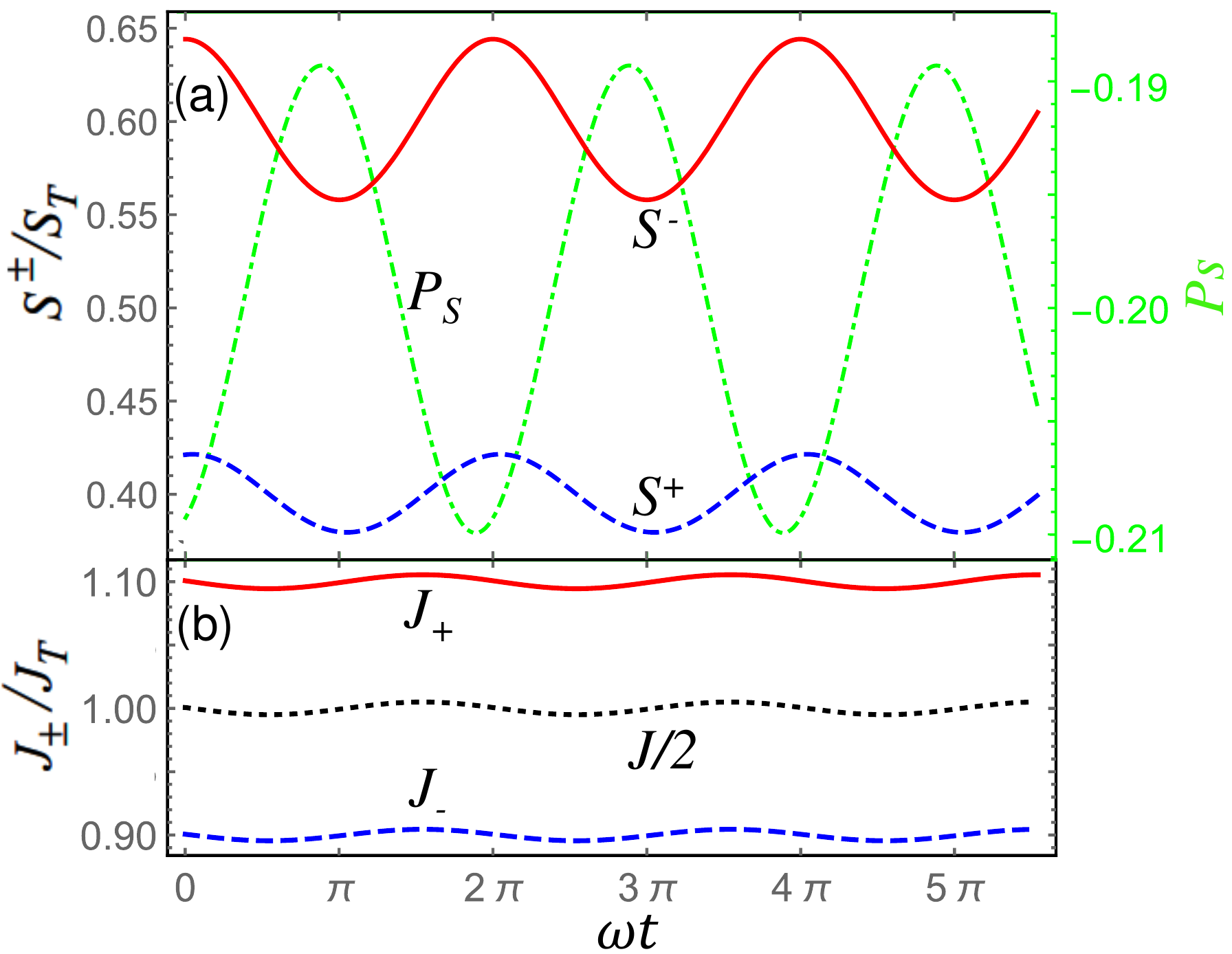}
\vspace{-0.5cm}
\caption{ Time evolution of (a) polarized light intensities $S^{\pm}$ and polarization $P_S$, (b) spin-polarized injection $J_{\pm}$ and their mean value $J/2$ under amplitude modulation. Light intensities are normalized to the intensity $S_T$ at twice the threshold $2J_T$. The injection $J_0=2 J_T$ with a constant spin polarization $P_J=0.1$, and the modulation frequency and amplitude are $\omega/2 \pi = 12$ GHz and $\delta J =0.01 J_T$, respectively. }
\label{fig:Evolution}
\vspace{-0.3cm}
\end{figure}

Surprisingly, the dynamic response of the circularly polarized  light in a spin-laser under AM has been largely overlooked, 
while a static regime was briefly considered~\cite{Basu2009:PRL}. In fact, a modulation of the total amplitude of spin-polarized injection changes both intensities of light helicities
(left and right circularly polarized light), as well as their relative ratio, which leads to a time-varying circular polarization of the laser emission, shown in Fig.~\ref{fig:Evolution}. 
With the current focus on adding spin-polarized carriers to vertical cavity surface emitting lasers (VCSELs)~\cite{Michalzik:2013}, employing experimentally simpler AM 
could be an important step towards realizing their dynamical operation at room temperature with electrical spin injection.
This breakthrough could overcome the present limitation of a high-speed and low-power operation constrained to optically-injected spin-lasers~\cite{Lindemann2019:N}.

Motivated by this situation, we investigate a nontrivial polarization response of a spin-laser under AM 
and explore the potential of such a response in digital communication. We use the rate equations for 
spin-lasers~\cite{Rudolph2003:APL,Rudolph2005:APL,Holub2007:PRL,Gothgen2008:APL,Lee2010:APL,Lee2012:PRB, Lee2014:APL},
which can be expressed in terms of spin- or helicity-resolved quantities~\cite{Gothgen2008:APL, Lee2010:APL},
\begin{eqnarray}
dn_{\pm}/dt &=& J_{\pm}-g_{\pm}S^{\mp}-(n_{\pm}-n_{\mp})/\tau_s-R_{sp}^{\pm}, 
\label{eq:ren}
\\
dS^{\pm}/dt &=& \Gamma g_{\mp}S^{\pm}-S^{\pm}/\tau_{ph}+\beta \Gamma R_{sp}^{\mp}, 
\label{eq:reS}
\end{eqnarray}
\noindent{where} $n_{\pm}$ are the densities of spin-up (down) $+$ ($-$) electrons (total $n=n_++n_-$), with a spin-relaxation time $\tau_s$. 
Since holes typically have much shorter spin-relaxation times than those of electrons
~\cite{Zutic2004:RMP}, only the electrons are spin polarized, while charge neutrality yields the densities of holes $p_{\pm}=(n_+ + n_-)/2$. $J_{\pm}$, where $J=J_+ + J_-$, are the injection rates of spin-up (down) $+$ ($-$) electrons. $S^{\pm}$, where $S=S^+ + S^-$, are the photon densities of positive ($+$) and negative ($-$) helicity.  
We also introduce the  polarization of the injection, $P_J=(J_+ - J_-)/J$, and the emitted light, $P_S=(S^+ - S^-)/S$.
The spontaneous recombination of carriers can be expressed in a linear form as $R_{sp}^{\pm}=n_{\pm}/\tau_r$, characterized by a carrier recombination 
time $\tau_r$~\cite{Bourdon2002:JAP,Hader2005:SPIE,Gothgen2008:APL}. 
The spin-dependent optical gain, $g_\pm$ can have various dependence on the carrier density~\cite{Xu2021:APL}.  In a linear model, 
it takes the form $g_{\pm}= g_0 (n_{\pm} + p_{\pm}-n_\text{tran})$~\cite{Gothgen2008:APL}, with $g_0$ the gain parameter and $n_\text{tran}$ the transparency carrier density. $\Gamma$ is the optical confinement factor and $\tau_{ph}$ is the photon lifetime in the cavity. The spontaneous emission factor $\beta~\sim 10^{-5}-10^{-3}$~\cite{Rudolph2005:APL,Holub2007:PRL},
characterizes the fraction of spontaneous emission coupled to the lasing mode. 

We first consider  a small-signal analysis (SSA) for a spin-laser under AM. 
The resulting decomposition $X(t)=X_0 +\delta X(t)$, into a steady-state and a small modulated part, is applied for $J_{\pm}$, $n_{\pm}$ and $S^{\pm}$. 
The modulated quantities take the form $\delta X(t)=Re [ \delta X_\omega e^{i \omega t}] $,
where $\omega$ is the angular modulation frequency. 
Under AM, $J=J_0 + \delta J \cos \omega t$, with
$P_J=P_{J0}$,  such that $J_{\pm} = (1 \pm P_J) J /2$. 
Due to the simultaneous existence of 
$n_{\pm} $ and $ p_{\pm}$, 
an analytical solution of SSA is very lengthy. 

To retain the analytical understanding, which still captures the main trends, we consider a simplified linear gain model
$g_{\pm}=g_0 (n_{\pm} -n_\text{tran}/2)$.
In the limit of a long spin-relaxation time, $\tau_s \gg \tau_r$,  
we obtain 
\begin{eqnarray}
\delta S^{\pm}_\omega = \frac{(g_0 S_0^{\pm}+ \beta/\tau_r )\delta J_{\mp}} {\frac{W_{\mp}}{\tau_r} +\frac{\beta G_0^{\mp}} {\tau_r} +  g_0 G_0^{\mp} S_0^{\pm} +g_0 S_0^{\pm} W_{\mp} -i \omega W_{\mp}  }, 
\end{eqnarray}
where, we define
$G_0^{\pm} \equiv g_0 (n_{0\pm} - n_\text{tran}/2)$ and $W_{\pm} \equiv \left( - i \omega -\Gamma G_0^{\pm} +1/\tau_{ph} \right)/\Gamma$.
Therefore, the resonance frequencies, $\omega_R^{\pm}$,  for $\delta S^{\pm}$, up to the linear order in $\beta$, are
\begin{eqnarray}
{\omega_R^{\pm}}^2 \approx  \frac{g_0 S_0^{\pm}} {\tau_{ph}} - \frac{ (\frac{1}{\tau_r} +g_0 S_0^{\pm} )^2} {2}  -\frac{\beta}{\tau_r} (\Gamma   g_0  n_{0\mp} -\frac{1}{\tau_{ph}}),
\label{eq:omegaR}
\end{eqnarray}
where the steady-state relation $1/\tau_{ph}- \Gamma G_0^{\pm}= \Gamma \beta n_{0\pm}/(S_0^{\mp} \tau_r)$ has been used. 
For $\beta \to 0$, we obtain
\begin{eqnarray}
\delta S^\pm_\omega \approx \frac{\Gamma g_0 S_0^{\pm} \delta J_{\mp}} { -\omega^2 -i \omega (1/\tau_r +g_0 S_0^{\pm} ) +g_0 S_0^{\pm} /\tau_{ph}  },
\end{eqnarray}
which, for $P_J=0$,  reduces to the result from conventional lasers~\cite{Chuang:2009}. 
The corresponding resonance frequencies become $\omega_R^{\pm} = \sqrt { g_0 S_0^{\pm}/\tau_{ph}   -(1/\tau_r +g_0 S_0^{\pm} )^2/2}$.

We next analyze the time evolution of the circular polarization of the emitted light, $P_S$. 
We denote the complex amplitude as $\delta S^{\pm}_\omega= |\delta S^{\pm} _\omega| e^{i\phi_{\pm}} $, where 
\begin{flalign}
& |\delta S^{\pm}_\omega | = \frac{\Gamma g_0 S_0^{\pm} \delta J_{\mp}  } { \sqrt {\left( \omega^2 -  g_0 S_0^{\pm}/\tau_{ph} \right)^2 + \omega^2 (1/\tau_r +g_0 S_0^{\pm} )^2}}, 
\label{eq:deltaS}
\\
& \tan \phi_{\pm} = \omega (1/\tau_r +g_0 S_0^{\pm} )/(-\omega^2 +  g_0 S_0^{\pm}/\tau_{ph}). 
\label{eq:phi}
\end{flalign}
$P_S$ can be generally expressed as 
\begin{eqnarray}
P_S (t) = \frac{ S_0^+ - S_0^- + \delta S^+ (t) -  \delta S^- (t) }{ S_0^+ +S_0^- +   \delta S^+ (t)+  \delta S^- (t)},
\end{eqnarray}
where $\delta S^{\pm} (t) = |\delta S^\pm_\omega|\cos (\omega t +\phi_\pm) $, or decomposed 
as $P_S (t) =P_S^0 +  \delta P_S (t)$, where $P_S^0= (S_0^+ - S_0^-)/S_0$, with $S_0=S_0^+ +S_0^-$. 
For SSA, $|\delta S^{\pm}_\omega| \ll S_0$, therefore 
\begin{eqnarray}
\delta P_S (t) \approx \frac{  |\delta S^+ _\omega|\cos (\omega t +\phi_+)  - |\delta S^-_\omega| \cos (\omega t +\phi_-)  }{ S_0 }. 
\label{eq:PS}
\end{eqnarray}
This AM response describes the results from Fig.~\ref{fig:Evolution} with $P_J=0.1$, 
$\omega/2 \pi = 12$ GHz, and $\delta J =0.01 J_T$.  Unless otherwise specified, the parameters in our calculations are guided by 
the fabricated spin-VCSELs~\cite{Rudolph2005:APL,Holub2007:PRL,Lee2010:APL,Lee2014:APL}$:  \Gamma=0.029$, 
$n_\text{tran}=4.0 \times 10^{17}$ cm$^{-3}$, $\tau_r=200$ ps, $\tau_{ph}= 1.0$ ps, $\tau_s=200$ ps, and $g_0=1.0 \times 10^{-5}$ cm$^3$s$^{-1}$.
$S^{\pm}$ undergo sinusoidal oscillations with separate mean values and the variation of $S^+/S^-$ due to $P_J\neq 0$, 
which leads to oscillations, $\delta P_S (t)=A \cos(\omega t + \varphi)$, around $P_{S0}=-0.2$. In conventional lasers, $P_J=0$ implies $P_S=0$. 
The polarization oscillation amplitude of the emitted light is
\begin{eqnarray}
A=\sqrt{A_+^2 +A_-^2 + 2A_+ A_- \cos(\phi_+- \phi_-)},
\label{eq:POamplitude}
\end{eqnarray}
$A_{\pm} = |\delta S^{\pm}_\omega|/S_0$ gives the superposition of two harmonic oscillations,  
$S_0^+ \neq S_0^-$, $|\delta S^+_\omega | \neq |\delta S^-_\omega |$ and $ \phi_+ \neq \phi_-$.

\begin{figure}[b]
\centering
\includegraphics*[width=8.6cm]{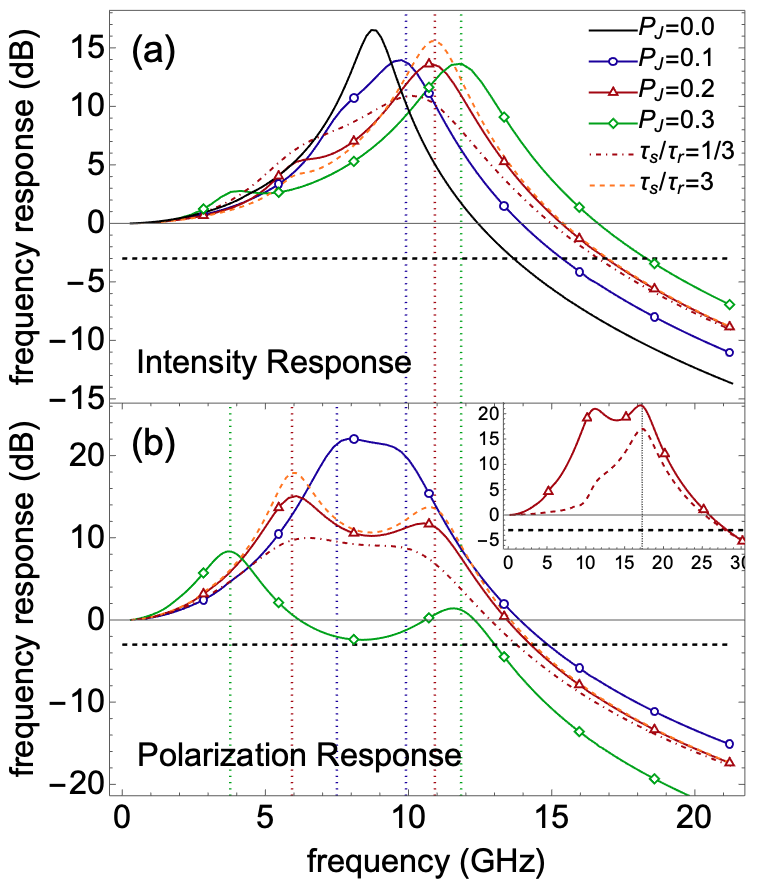}
\vspace{-0.5cm}
\caption{AM: (a) Intensity and (b) polarization response.  $P_J=0.1$, $0.2$, and $0.3$. $\tau_s/\tau_r=1$, except for  
$\tau_s/\tau_r=1/3$ and $3$ with $P_J=0.2$. 
$J_0=3 J_T$, $\beta=10^{-4}$. The simplified model $g_{\pm}=g_0 (n_{\pm} -n_\text{tran}/2)$ is used.  
Vertical dashed lines: 
the resonance frequencies given by Eq.~(\ref{eq:omegaR}). Horizontal dashed lines: 
the -3 dB response, which indicate the modulation bandwidth. Inset: 
the polarization (solid) and intensity (dashed) response with the full linear gain model for $P_J=0.2$. }
\label{fig:Response}
\vspace{-0.9cm}
\end{figure}

This picture of the two harmonic oscillations also characterizes the AM results in Fig.~\ref{fig:Response} for  the response of $S^-$ (majority spin) and $P_S$. 
The intensity response function, $R(f)=\delta S^-/\delta J_+$~\cite{Lee2010:APL}, for $P_J \to 0$ reduces to the usual form, $R(f)=\delta S/\delta J$, in conventional lasers~\cite{Chuang:2009}.
In contrast, the polarization response function, $R(f)=\delta P_S/\delta J$, has no conventional counterpart.  Here $\delta S^-$ and $\delta P_S$ are 
the amplitudes of the intensity and polarization responses. These responses are normalized to their low-frequency values, $f_{\rm low}$, as $\bar R(f) = 10 \log_{10} [R^2(f)/R^2(f_{\rm low}) ]$. 

While the intensity response peak in Fig.~\ref{fig:Response}(a)   
is near $\omega=\omega_R^-$, given by Eq.~(\ref{eq:omegaR}), and shows an enhanced $\omega_R^-$ with $P_J$~\cite{Lee2010:APL},  
the polarization response in Fig.~\ref{fig:Response}(b) reveals the presence of two peaks. This can be understood because the amplitude of $P_S$ oscillations involves contribution from both 
$A_{\pm}$, as given in Eq.~\ref{eq:POamplitude}, with unequal resonance frequencies,  $\omega_R^+ \neq \omega_R^-$. 
Near $\omega_R^+$ ($\omega_R^-$), the amplitude $A_+$ ($A_-$) reaches its maximum value, which results in the lower (higher) peak of a frequency response 
in Fig.~\ref{fig:Response}(b). 
$\omega_R^{\pm}$ from 
Eq.~(\ref{eq:omegaR}) are marked by vertical dashed lines, which coincide with the peaks of the intensity and polarization responses, 
showing a good agreement between analytical and numerical results, in which we also use a simplified gain model, $g_{\pm}= g_0 (n_{\pm} -n_\text{tran}/2)$. However, this obtained agreement
and the main trends in the frequency responses are not limited to the simplified gain model. Even with a more general model, $g_{\pm}= g_0 (n_{\pm} + p_{\pm}-n_\text{tran})$,
the results from the inset of Fig.~\ref{fig:Response}(b) for $P_J=0.2$  confirm a similar behavior in the frequency responses, only with slightly shifted resonant frequencies, justifying
our use of a simplified gain model.

\begin{figure}[b]
\centering
\includegraphics*[width=8.6cm]{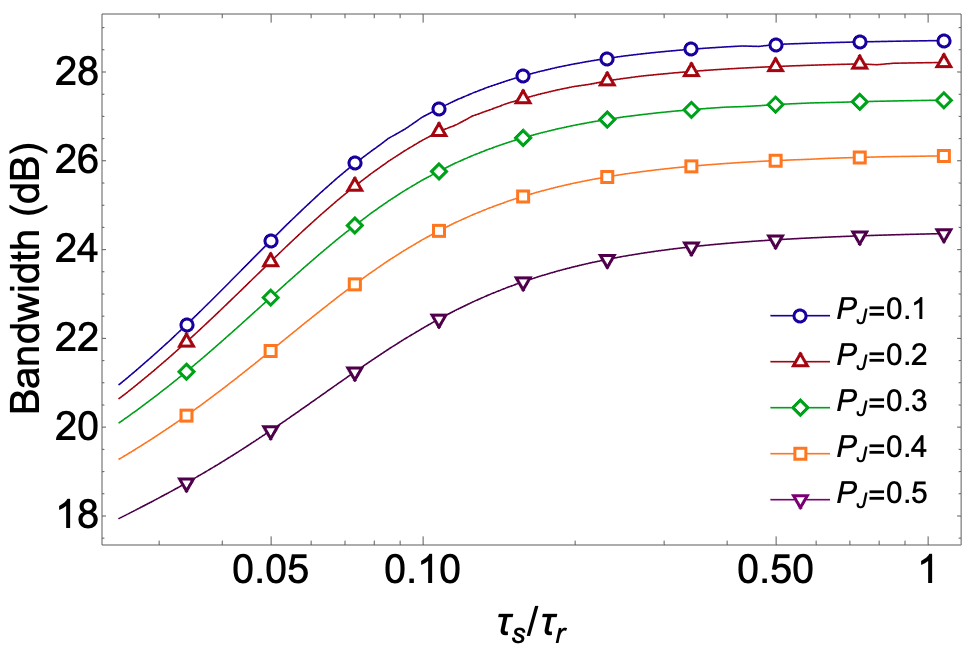}
\vspace{-0.5cm}
\caption{The bandwidth dependence of the polarization response under AM 
with the normalized spin-relaxation time, $\tau_s/\tau_r$, for the polarization of injection, $P_J=0.1$ to $0.5$.}
\label{fig:taus}
\vspace{-0.3cm}
\end{figure}

Since the electron spin-relaxation time, $\tau_s$, is a key quantity that affects polarization response, it is important to understand its role on the modulation bandwidth, $f_\text{3dB}$,  which represents a usable frequency range, with the frequency response above -3 dB.  $f_\text{3dB}$ is depicted by horizontal dashed lines in Fig.~\ref{fig:Response}. Therefore, 
we calculate the dependence of bandwidth on $\tau_s$, normalized by recombination time $\tau_r$. Even for the same
gain material, $\tau_s/\tau_r$ can strongly change with the growth direction~\cite{Iba2011:APL,Iba2020:PSPIE,Ohno2020:APE}, 
an applied gate voltage or magnetic field~\cite{Zutic2004:RMP}. 
From the steady-state regime and at a fixed $P_J$~\cite{Gothgen2008:APL}, we can see several factors to generalize our AM description, focused on the analysis of the gain region, which is more accurate for the optical spin injection~\cite{Rudolph2003:APL,Gerhardt2011:APL,Iba2011:APL,Frougier2013:APL,Alharthi2014:APL,Lindemann2019:N}, as electrical spin injection also involves spin-dependent transport from the injector. Both the lasing threshold reduction, as compared to the $P_J=0$ limit, and the spin amplification, in which modest polarization of the carriers in the gain region yields a much larger $P_S$~\cite{Iba2011:APL}, depend on the $\tau_s/\tau_r$ ratio~\cite{Gothgen2008:APL}. This implies several additional AM mechanisms 
for $\delta P_S$. With strong transport nonlinearities in semiconductors there is a bias-dependent carrier polarization and 
$P_S$~\cite{Zutic2004:RMP, Zutic2001:PRB,Zutic2002:PRB,Zutic2007:JPCM}.  An electrical spin injection in semiconductors and their light-emitting diodes can lead to both bias-dependent spin-relaxation time and the spin-dependent transport, for example, due to the barrier properties~\cite{Zutic2004:RMP,Giba2020:PRA,Barate2017:PRA,Crooker2009:PRB}. A change in the applied bias could produce additional contributions to $\delta P_S$, to be included in Eqs.~(\ref{eq:deltaS}) and~(\ref{eq:phi}) and used in  Eq.~(\ref{eq:PS}), or similar generalization beyond the simplified gain model.

We clearly see two trends in Fig.~\ref{fig:taus}: the bandwidth increases with $\tau_s$ and decreases with $P_J$. 
The second trend appears counterintuitive since: (i) in conventional lasers an increase in $f_R= \omega_R/(2\pi) $ is 
related to the increase in the bandwidth, as expressed by $f_\text{3dB}\approx \sqrt{1+\sqrt{2}} f_R$~\cite{Michalzik:2013}
and (ii) from Eq.~(\ref{eq:omegaR}), as shown in the early of the AM in spin-lasers~\cite{Lee2010:APL}, 
that $f^-_R \propto \sqrt{(1+|P_J|/2)(J_0/J_T-1)}$. While (i) and (ii) would suggest that from an increase in $f^-_R$ with $P_J$
we should expect an {\em increase} in $f_\text{3dB}$ with $P_J$, both Figs.~\ref{fig:Response}(b) and  \ref{fig:taus} reveal an opposite trend. 
Since $\omega_R^-$, corresponding to the helicity light from the majority spin,  becomes larger with $P_J$, while the $\omega_R^+$ becomes smaller, there is a larger frequency separation between $\omega_R^{\pm}$ and lower response values. Consequently, at a smaller $f$, the response drops below -3 dB and $f_\text{3dB}$ is reduced. However, it does not mean that the smallest $P_J$ would be ideal for a larger $f_\text{3dB}$. This is because the evaluation of $f_\text{3dB}$ includes a normalization. The actual value of $\delta P_S$ decreases with  $P_J$. For applications, $\delta P_S$ should be large enough to be measurable and have a sufficient signal-to-noise ratio~\cite{Lee2010:APL,Wang2021:OE,Li2020:APL}. Therefore, an optimization between  the magnitudes of $P_J$ and $f_\text{3dB}$ is needed. 

In addition to previously discussed analog operation, we can also examine the digital operation of spin-lasers, important for digital data transfer~\cite{Michalzik:2013,Agrawal:2002}.
Similar to conventional lasers, the corresponding AM is expressed as
\begin{equation}
J(t)=J_0+\delta J(t)=J_0+\delta J f(t),
\label{eq:digital}
\end{equation}
where $\delta J$ is the modulation amplitude and, unlike our previously considered harmonic dependence,  $f(t)$ is the binary function
of the coded input signal. A binary ``0" (``1") is set to $J_0$ ($J_0+\delta J$). In
the previous AM studies of conventional or spin-lasers, the emitted photon density would reflect the information encoded 
in the input $J(t)$~\cite{Michalzik:2013,Wasner2015:APL}. Bit ``0" (``1") is then defined by being below (above) some specified threshold in $S$. 
However, here we focus on the polarization of the emitted light, such that ``0" (``1") is defined with respect to a threshold in $P_S$.

To analyze the quality of a digital signal, we use the eye diagrams. The size of the central ``eye" opening indicates the distinguishability between digital ``0" and ``1" signals.
We simulate a binary signal, by using $2^{10}$ pseudorandom bits with a common non-return-to-zero modulation, the pulse remains on throughout the bit slot, and its amplitude does not fall to zero between successive ``1" bits. This stream of bits is first filtered by a generalized raised cosine filter~\cite{Agrawal:2002} to reduce parasitic ringing effects which complicate distinguishing ``0" and ``1." 

\begin{figure}[t]
\centering
\includegraphics*[width=8.6cm]{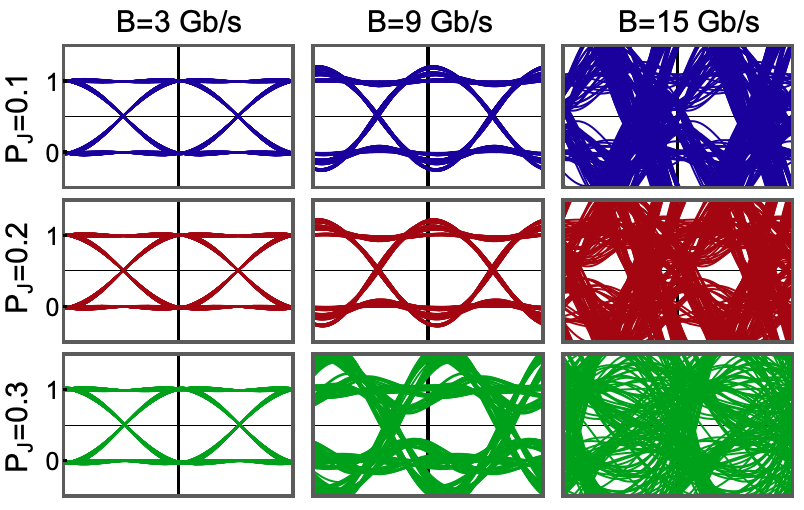}
\vspace{-0.5cm}
\caption{Eye diagrams of a normalized polarization of the emitted light under AM for injection with $P_J=0.1$, $0.2$, and $0.3$, for a series of bit rates $B=3, 9, 15$ Gb/s. Parameters: $J_0=3 J_T$, $\beta=10^{-4}$, $
\tau_s/\tau_r=1$. }
\label{fig:EyeDiagram}
\vspace{-0.3cm}
\end{figure}

The corresponding eye diagrams in Fig.~\ref{fig:EyeDiagram} are 
generated by dividing the laser wave form into segments of an equal size of two bits and overlaying them. The bit slot time is an inverse of a bit rate, $B$.
By separately changing these bit rates,  $B=3, 9, 15$ Gb/s, and the polarization injection, $P_J=0.1, 0.2, 0.3$, we can identify several trends for the digitally-encoded circular polarization under AM.
We see that for a small polarization, $P_J=0.1$, the data transfer with a bit rate up to 15 Gb/s remain efficient. As $P_J$ increases, the central eye in the diagrams starts to close, especially at large bit rates, indicating worse performance with $P_J$.

This trend in $P_J$ is consistent with the behavior illustrated in Figs.~\ref{fig:Response}(b) and \ref{fig:taus}, which both show a decreasing bandwidth with a larger $P_J$. 
We note that the bit rate of open eye diagrams is smaller than the bandwidth. This can be understood because the time evolution of $P_S$ depends not only on the separate amplitudes, $S^+$ 
and $S^-$, but also on the phase differences between $S^+$ and $S^-$, as shown by Eq.~\ref{eq:POamplitude}. From numerical calculations, we find that the phase difference depends on various factors including $\omega$, $B$,  
and $P_J$. 
In the eye diagrams, abrupt changes in the pseudorandom input with large bit rate leads to a transient effect~\cite{Boeris2012:APL}, which influences 
the time-variation of the phase difference and the shape of $P_S$. As a result, the maximum bit rate supporting open 
eye diagrams is suppressed, compared to the bandwidth in the response curves from Figs.~\ref{fig:Response}(b) and \ref{fig:taus}. 

With a simple amplitude modulation in spin-lasers, we have revealed overlooked nontrivial dynamics of their circularly polarized emitted light.
Several identified trends have been corroborated by a combination of analytical and numerical methods, within the small signal analysis for 
both harmonic and digital modulation. Our findings, based on the rate-equation model in Eqs.~(\ref{eq:ren}) and (\ref{eq:reS}), suggest several generalizations.
One can combine this phenomenological approach with a microscopic gain calculation~\cite{FariaJunior2015:PRB} and also include the hole spin-relaxation
time, which could be important for GaN-based lasers~\cite{FariaJunior2017:PRB}, but was not considered in their experimental analysis~\cite{Bhattacharya2017:PRL}.
Furthermore, it would be important to extend our findings to other types of rate-equation models which describe the effects of optical 
anisotropies~\cite{Gerhardt2011:APL,SanMiguel1995:PRA,Dyson2003:JOBQS,Adams2018:SST,Xu2021:PRB,Adams2022:IEEEJQE,Yokota2023:IEICE,Panajotov2023:P}, 
or the presence of external cavities~\cite{Frougier2013:APL,Alouini2018:OE}.

Currently, dynamical room-temperature operation of VCSELs is limited to the optical injection of spin-polarized carriers~\cite{Lindemann2019:N}. Using the amplitude modulation 
is promising for the push towards their room-temperature operation with electrical spin injection. This effort could incorporate advances in ferromagnetic contacts 
with a perpendicular magnetization to remove the need for an applied magnetic field~\cite{Zutic2020:SSC,Sinsarp2007:JJAP,Hovel2008:APLa,Liang2014:PRB,Giba2020:PRM,Tao2018:N,Cadiz2018:NL}. 
Beyond the usual III-V gain regions, with a growing family of van der Waals materials, it could also be possible to combine a
proposal for spin-lasers with an atomically-thin gain region~\cite{Lee2014:APL}, where the spin-polarized carriers would be provided
by electrically-tunable magnetic proximity effects~\cite{Zutic2019:MT,Liang2023:NE}.

For a full signal transduction between the carrier spin and the helicity of light, 
to enable versatile applications of spin-lasers for optical communication, high-performance interconnects, 
holographic information~\cite{Ni2022:NC}, or three-dimensional displays~\cite{Nishizawa2021:MM}, 
another focus should be on the development of helicity detectors. Even for simple ferromagnetic contacts with GaAs, the optimization of such detectors
requires a careful understanding of their dynamical properties~\cite{Safarov2022:PRL}.

\vspace{0.5cm}
We thank N. C. Gerhardt for valuable discussions.
This work has been supported by the National Natural Science Foundation of China (Grant No. 12104118), the NSF ECCS-2130845, and AFOSR FA9550-22-1-0349.

\section{Author Declarations}

\subsection{Conflict of interest}
The authors have no conflicts to disclose. 

\section{Data Availability}
The data that supports the findings of this study are available within the article. 

\end{document}